\documentclass[%
 notitlepage, 10pt, twocolumn,
 amsmath,amssymb,
 aps,prx,
floatfix,
]{revtex4-2}

\usepackage[dvipsnames]{xcolor}
\usepackage{hyperref}
\hypersetup{
  breaklinks=true,
  colorlinks=true,
  allcolors=BlueViolet,
}
\usepackage{tikz}
\usepackage{graphicx}

\usepackage{physics}
\usepackage{amsthm}
\usepackage{graphicx}
\graphicspath{{./figures/}} 
\usepackage{dcolumn}
\usepackage{hyperref}
\usepackage[linesnumbered,boxed]{algorithm2e}
\usepackage{mathtools}
\usepackage{xcolor}
\usepackage{subcaption}
\usepackage{caption}
\usepackage[percent]{overpic}

\begin{document}

\preprint{}
\title{Enhanced quantum sensing mediated by a cavity in open systems}

\author{Quinn Langfitt$^1$}
\email{quinn.langfitt@gmail.com}

\author{Zain H. Saleem$^1$}
\email{zsaleem@anl.gov}

\author{Michael Perlin$^2$}
\email{mika.perlin@gmail.com}

\author{Tian Zhong$^3$}
\email{tzh@uchicago.edu}

\author{Anil Shaji$^4$}
\email{shaji@iisertvm.ac.in}

\author{Stephen K. Gray$^5$}
\email{gray@anl.gov}

\affiliation{Mathematics and Computer Science Division, 
Argonne National Laboratory, 9700 S Cass Ave, Lemont IL 60439, USA $^1$}

\affiliation{Infleqtion, Inc., Chicago, Illinois 60615, USA $^2$}

\affiliation{Pritzker School of Molecular Engineering, University of Chicago, Chicago, Illinois 60637, USA $^3$}

\affiliation{School of Physics, IISER Thiruvananthapuram, Kerala, India, 695551 $^4$}
\affiliation{Center for Nanoscale Materials,
Argonne National Laboratory, Lemont, Illinois 60439, USA $^5$}


\begin{abstract}
We simulate the dynamics of systems with multiple qubits coupled to a cavity in order to assess
their potential for quantum metrology of a parameter in the open systems limit. The qubits and
the cavity are both allowed to have losses and the system is studied under various coupling strength
regimes. The focus is primarily on the coupling between the qubits using the quantum Fisher
information as the measured parameter. Some results on estimating the qubit-cavity detuning
parameter are also presented. We investigate the scaling of the uncertainty in the estimate of
the qubit-cavity coupling with the number of qubits and for different initial states of the qubits
that act as the quantum probe. As initial probe states, we consider Dicke states with varying
excitation numbers, the GHZ state, and separable X-polarized states. It is shown that in the strong
coupling regime, i.e., when the coupling between the qubits and the cavity is greater than the
decay parameters of both the qubits and the cavity, Dicke states with a large excitation number
can achieve the Heisenberg limit, with the precision scaling improving as the excitation number
increases. A particularly intriguing finding of our study is that in the weak coupling regime, as well
as in situations where either the qubit or cavity decay parameters exceed the coupling, the separable
X-polarized state is the best in terms of scaling and exhibits strongly superlinear,
and in some regimes near-quadratic, finite-size QFI scaling over the system sizes considered. 
\end{abstract}

\maketitle


\section{Introduction}

In quantum metrology, the goal is to estimate parameters more precisely using quantum resources, such as entanglement and squeezing \cite{braunstein1994statistical, giovannetti2004quantum,toth2014quantum,nawrocki2015introduction,polino2020photonic}. The maximum achievable precision is set by the Heisenberg Limit (HL), which corresponds to the precision scaling as $1/N$, where $N$ is the number of elementary sub-units making up the quantum probe. HL can be achieved with entangled states in a noiseless environment \cite{giovannetti2004quantum}. However, system noise can hinder the probe's capabilities, leading to a precision scaling of $1/\sqrt{N}$, referred to as the Standard Quantum Limit (SQL) or worse. Our research aims to surpass this limitation through the careful selection of probe states. For instance, it has been demonstrated that using a Dicke state with certain excitation numbers as the probe, under the Tavis-Cummings model in the strong coupling regime, can exceed SQL and reach HL, when the duration of the measurement is optimised, owing to the Dicke state's resilience to noise \cite{saleem2023achieving}.

In this study, we focus on the $N$ qubits-cavity model, limiting ourselves to permutationally symmetric states. Permutational symmetry reduces the Hamiltonian's dimension size to $O(N^5)$, enabling us to extend simulations to a larger number of qubits than otherwise possible \cite{chase2008collective, shammah2018open}. The $N$ qubits are coupled to a cavity with a coupling strength of $g$, with decay rates $\kappa$ and $\gamma$ for the resonator and spin states, respectively. We normalize these decay rates with respect to $g$. Moreover, aligning with our focus on quantum sensing in noisy environments, we primarily concentrate on high-loss regimes — noisy-cavity, noisy-qubit, and weak coupling — where either $\kappa/g$ or $\gamma/g$, or both, are significantly greater than one, and compare these to the less noisy, strong coupling regime. A noticeable difference in the selection of optimal probe states is observed when transitioning from the strong-coupling to high-loss regimes. Fig.~1 illustrates these four regimes.
\begin{figure}[!htb]
    \centering
    \resizebox{0.8\columnwidth}{0.8\columnwidth}{%
    \begin{tikzpicture}
        \fill[gray!15] (0,0) rectangle (2,2); 
        \fill[blue!15] (2,0) rectangle (4,2); 
        \fill[blue!15] (0,2) rectangle (2,4); 
        \fill[blue!15] (2,2) rectangle (4,4); 

        \draw (0,0) rectangle (4,4);
        \draw (2,0) -- (2,4);
        \draw (0,2) -- (4,2);

        \node[font=\tiny] at (1,1) {Strong Coupling};
        \node[font=\tiny] at (3,1) {Noisy-Cavity};
        \node[font=\tiny] at (1,3) {Noisy-Qubit};
        \node[font=\tiny] at (3,3) {Weak Coupling};

        \node[font=\small] at (2,-0.8) {$\kappa/g$};
        \node[font=\small] at (-1,2) {$\gamma/g$};

        \node[font=\small] at (2,-0.3) {$1$}; 
        \node[font=\small] at (4,-0.3) {$3$}; 
        \node[font=\small] at (-0.3,2) {$1$}; 
        \node[font=\small] at (-0.3,4) {$3$}; 

        \node[font=\small] at (-0.2,-0.2) {$0$}; 

        \draw[->] (0,0) -- (4.5,0) node[anchor=north] {}; 
        \draw[->] (0,0) -- (0,4.5) node[anchor=east] {}; 
    \end{tikzpicture}
    }
    \caption{Coupling regimes. The boxes colored in blue are the main focus of this paper.}
    \label{fig:coupling_regimes}
\end{figure}

The quantum Fisher information (QFI) serves as our tool for measuring the accuracy of our probe states \cite{Faist:21, Akipour2014Quantum, Naghiloo2019, lu2010quantum, gammelmark2014fisher, altintas2016quantum}. The QFI quantifies the sensitivity of a quantum state to changes in the parameter being estimated, which is inversely proportional to the variance of the estimated parameter (see Appendix \ref{app:qfi} for further detail). 
Due to the open evolution of our quantum probe, the QFI has a non-trivial time dependence.  For closed systems, the QFI increases quadratically over time, but for open systems, there is a time for which the QFI is maximized and the optimal time for measurement is achieved \cite{saleem2022optimal}. In our study, we consider the maximum QFI over time, and then see how the maximum QFI value scales with increasing system size $N$. 

The remainder of the paper is organized as follows: Section \ref{sec:model} describes in further detail the system under study. Section \ref{sec:max_QFI} examines a particularly successful state for high-loss regimes, the polarized-$X$ state, and its max QFI scaling across $N$. Section \ref{sec:comp_states} compares the results of the probe states across all four regimes, as depicted in Fig.~\ref{fig:coupling_regimes}. Section \ref{sec:detune_est} briefly discusses our results when estimating detuning $\Delta$. Finally, Section \ref{sec:conclusion} provides a conclusion and discussion of future work.

\section{N qubits-cavity system}
\label{sec:model}

In our study, we consider a system composed of $N$ two-level quantum systems (qubits) interacting with a single electromagnetic cavity/resonator mode. We include decays for both the resonator and the qubits, denoted as $\kappa$ and $\gamma$, respectively. These decay rates appear in the equation that governs the evolution of the open quantum system, which is a Gorini-Kossakowski-Sudarshan-Lindblad (GKSL) type master equation \cite{gorini_completely_1976, lindblad_generators_1976, chruscinski_brief_2017, manzano2020short}:
\begin{equation}
\label{eq:Lindblad}
    \dot{\rho} = -i [H, \rho] + \kappa \mathcal{D}[a](\rho) + \gamma \sum_{j=1}^N \mathcal{D}[\sigma_-^{(j)}](\rho),
\end{equation}
where \( \mathcal{D}[O](\rho) = O \rho O^\dagger - \frac{1}{2} \{O^\dagger O, \rho\} \) is the Lindblad dissipator for an operator \( O \). Eq.~\eqref{eq:Lindblad} encompasses both the unitary evolution driven by the system Hamiltonian and the incoherent processes due to environmental interactions. The system Hamiltonian, $H$ in Eq.~\eqref{eq:Lindblad} is given by,
\begin{equation}
		H = \frac{\omega_q}{2} \sum_{j=1}^N \sigma_z^{(j)} + \omega_c a^\dagger a + g \sum_{j=1}^N \big( a^\dagger \sigma_-^{(j)} + a \sigma_+^{(j)} \big). 
\end{equation}
where $\omega_c$ and $\omega_q$ are the frequencies of the cavity and the qubits, $a^{\dagger}$ and $a$ are the photon creation and annihilation operators for the cavity, $\sigma_z^{(j)}$ is the Pauli Z operator acting on the $j^{\rm th}$ qubit, $g$ is the coupling strength between the qubits and the cavity, and $\sigma_-^{(j)}$ and $\sigma_+^{(j)}$ represent the lowering and raising operators acting on the $j^{\rm th}$ qubit. 

We operate under resonance conditions and within a rotating frame, setting $\omega_q = \omega_c = 0$. Additionally, we consider the scenario of estimating detuning, characterized by a nonzero difference between $\omega_q$ and $\omega_c$, as an extra parameter of interest. In the case of detuning estimation, we set this difference $\Delta = 0.1 g$, ensuring that the system remains near-resonance and thereby avoiding significant alterations to the model under investigation.

Within the $N$ qubits-cavity system, we examine the metrology performance of the following probe states: the GHZ state, the Dicke states, and the X-polarized state, each representing different levels of entanglement. The GHZ state \cite{Greenberger1989} is a maximally entangled state expressed as 
\begin{equation}
    \ket{\text{GHZ}} = \frac{1}{\sqrt{2}} \big(\ket{0}^{\otimes N} + \ket{1}^{\otimes N}\big).
\end{equation}
The X-polarized state, so-named because it is an eigenstate of the sum of $\sigma_x$ over the individual qubits, with eigenvalue $N$, is given by 
\begin{equation}
    \ket{\text{X}} = \frac{1}{\sqrt{2^N}}(\ket{0} + \ket{1})^{\otimes N}.
\end{equation}
The Dicke states, containing intermediate levels of entanglement, are defined as superpositions of all permutations of $N$ qubits with a fixed number of excitations,
\begin{equation}
    \ket{\text{Dicke-n}} = \frac{1}{\sqrt{\binom{N}{n}}} \sum_{\text{permutations}} |\underbrace{0 \ldots 0}_{N-n}\underbrace{1 \ldots 1}_{n} \rangle.
\end{equation}

\begin{figure}[!htb]
    \centering
    \includegraphics[width=\columnwidth]{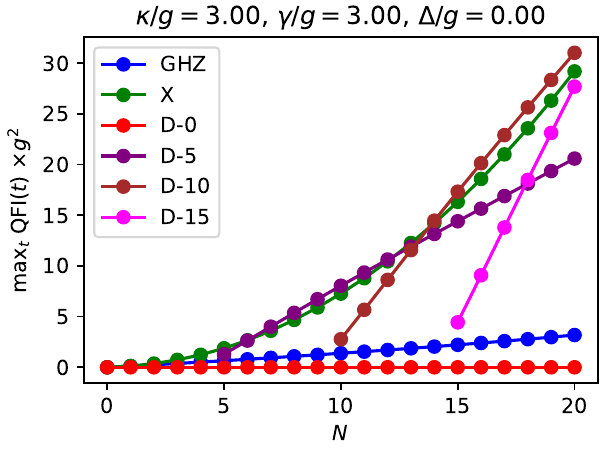}
    \caption{Scaling of maximum QFI with respect to number of qubits $N$}
    \label{fig:size_scaling_weak}
\end{figure}


\section{Scaling of maximum QFI}
\label{sec:max_QFI}

Throughout this work, the QFI is calculated from the full joint
qubit--cavity density matrix $\rho$, rather than from a reduced density
matrix of either the qubits or the cavity alone. The resulting QFI
therefore quantifies the maximum information about the estimated parameter
that is, in principle, accessible through an optimal measurement on the
joint qubit--cavity system. We do not restrict the measurement to a
particular observable or to measurements performed solely on the qubits
or solely on the cavity.

Figure~\ref{fig:size_scaling_weak} compares the scaling of the maximum QFI as a function of the number of qubits $N$ across Dicke states, GHZ state, and the $X$-state. We see that though the $X$-state has smaller maximum QFI than the Dicke-$10$ state at $N=20$, the $X$-state shows quadratic scaling, while the Dicke states appear to have linear scaling with respect to $N$. The Dicke-$0$ state, which is another separable state, appears to perform quite poorly. The GHZ state also seems to have scaling greater than linear with $N$.

\begin{figure}[!htb]
    \centering
    \begin{overpic}[width=\columnwidth]
        {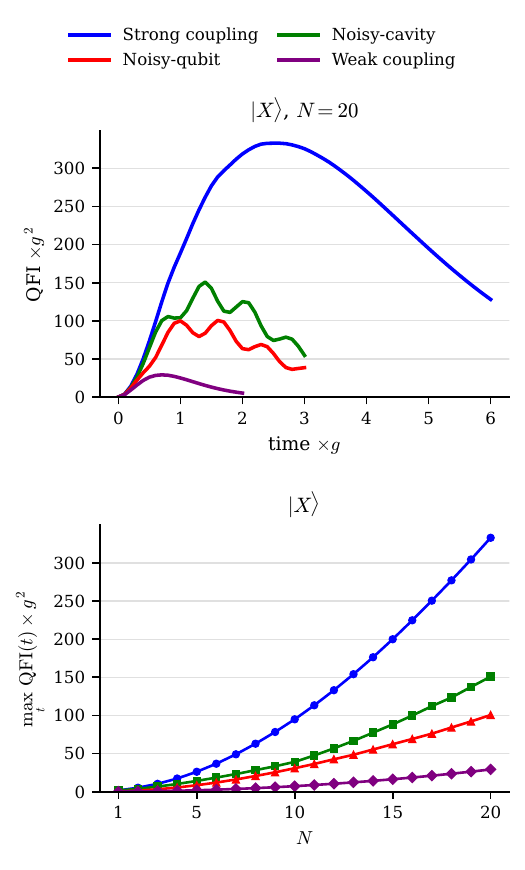}
        \put(4,84.5){\bfseries (a)}
        \put(4,39){\bfseries (b)}
    \end{overpic}
    \caption{
        Analysis of the $X$-state in various coupling regimes.
        (a) Time-dependent QFI for $N=20$.
        (b) Size dependence of the maximum QFI.
        The regimes are strong coupling
        ($\kappa/g=\gamma/g=0.8$),
        noisy-cavity
        ($\kappa/g=3.0$, $\gamma/g=0.2$),
        noisy-qubit
        ($\kappa/g=0.2$, $\gamma/g=3.0$),
        and weak coupling
        ($\kappa/g=\gamma/g=3.0$).
    }
    \label{fig:combined_figures}
\end{figure}

Figure~\ref{fig:combined_figures} illustrates the time-dependent behavior of QFI for the $X$-polarized state, under various coupling regimes and sets of decay rates $\kappa/g$ and $\gamma/g$, while keeping the number of qubits $N$ constant. The figure illustrates a consistent pattern where the QFI first increases, peaks, and then gradually diminishes. This trend is marked by oscillations, particularly pronounced in regimes of poor cavity and qubit quality. Notably, the regions exhibiting the highest QFI values and the corresponding maximal QFI both escalate in size with an increase in $N$ for all the probe states we consider, including the Dicke-$n$ states and the GHZ state \cite{saleem2023achieving}. 

%

Our study focuses on the scaling of the maximum QFI in high-loss regimes.
To characterize its dependence on system size over the numerically
accessible range, we fit the maximum QFI to
$\max_t F_Q(N)=aN^b+c$, where $b$ denotes the finite-size scaling
exponent. For reference, linear QFI scaling, $F_Q\propto N$, corresponds
to the standard quantum limit (SQL), whereas quadratic scaling,
$F_Q\propto N^2$, is conventionally associated with Heisenberg scaling.
We use these benchmarks to characterize the observed finite-size behavior,
without assuming that the fitted exponent determines the asymptotic
$N\rightarrow\infty$ scaling.

It is important to note that the parameter estimated here is the
qubit--cavity coupling $g$, which multiplies the collective interaction
generator
\[
G=\sum_{j=1}^{N}\left(a^\dagger\sigma_-^{(j)}
+a\sigma_+^{(j)}\right)
=a^\dagger J_-+aJ_+ .
\]
Although $G$ contains only a single sum over the qubits, its variance
contains cross terms between different qubits. For example, in the
unitary short-time limit, $F_Q(g)=4t^2\mathrm{Var}(G)$. For the initial
X-polarized state $|X\rangle=|+\rangle^{\otimes N}$ and the cavity
initially in the vacuum state, $\langle G\rangle=0$ and
\[
\langle G^2\rangle
=\langle J_+J_-\rangle
=\frac{N}{2}+\frac{N(N-1)}{4}
=\frac{N^2+N}{4},
\]
giving
\[
F_Q(g)=t^2(N^2+N).
\]
Thus, a quadratic contribution to the QFI can arise from coherent
cross terms associated with the common cavity mode. Accordingly, the
near-quadratic finite-size scaling observed here should be interpreted
as a cavity-mediated collective enhancement of sensitivity to $g$,
rather than, by itself, as evidence that an initially entangled probe
resource is responsible for the enhancement.

\begin{figure}
    \includegraphics[width=0.95\columnwidth]{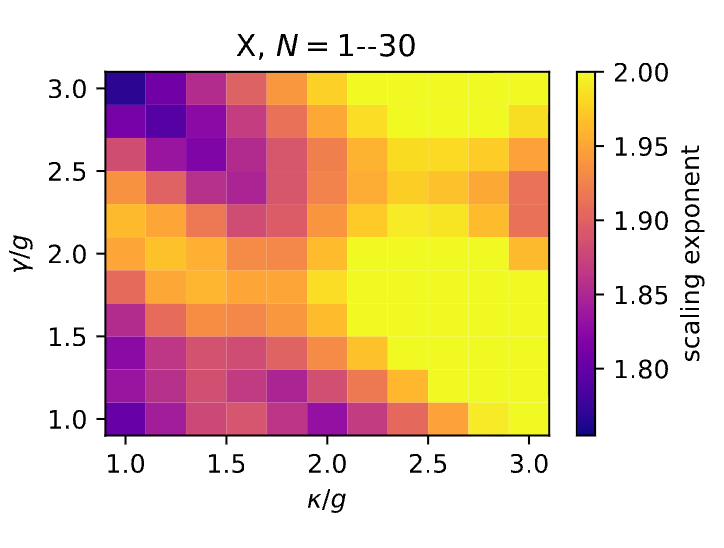}
    \caption{Finite-size scaling exponent $b$ for the X-polarized state as a function of the cavity and qubit decay rates $\kappa/g$ and $\gamma/g$. The exponent is obtained by fitting $\max_t F_Q(N)=aN^b+c$ over $N=1\text{--}30$, with $0\leq b\leq2$.}
    \label{fig:map_plot}
\end{figure}

In Fig.~\ref{fig:map_plot}, we show the finite-size scaling exponent $b$ for the X-state
over the full range of cavity and qubit decay rates considered. To examine
the robustness of the observed scaling behavior with increasing system size,
we extend the X-state calculations to $N=30$ and fit the maximum QFI according
to $\max_t F_Q(N)=aN^b+c$ over $N=1$--$30$. The fitted exponent remains
strongly superlinear throughout the parameter region considered, with
$b \approx 1.8$--$2.0$. Thus, the enhanced scaling observed for $N\leq 20$
persists over the extended system-size range. We emphasize, however, that
$b$ is a finite-size scaling exponent over the numerically accessible range
and should not be interpreted as establishing the asymptotic
$N\rightarrow\infty$ scaling.

\section{Comparison of Probe States}
\label{sec:comp_states}

\begin{figure*}[t]
    \centering
    \begin{overpic}[width=\textwidth]
        {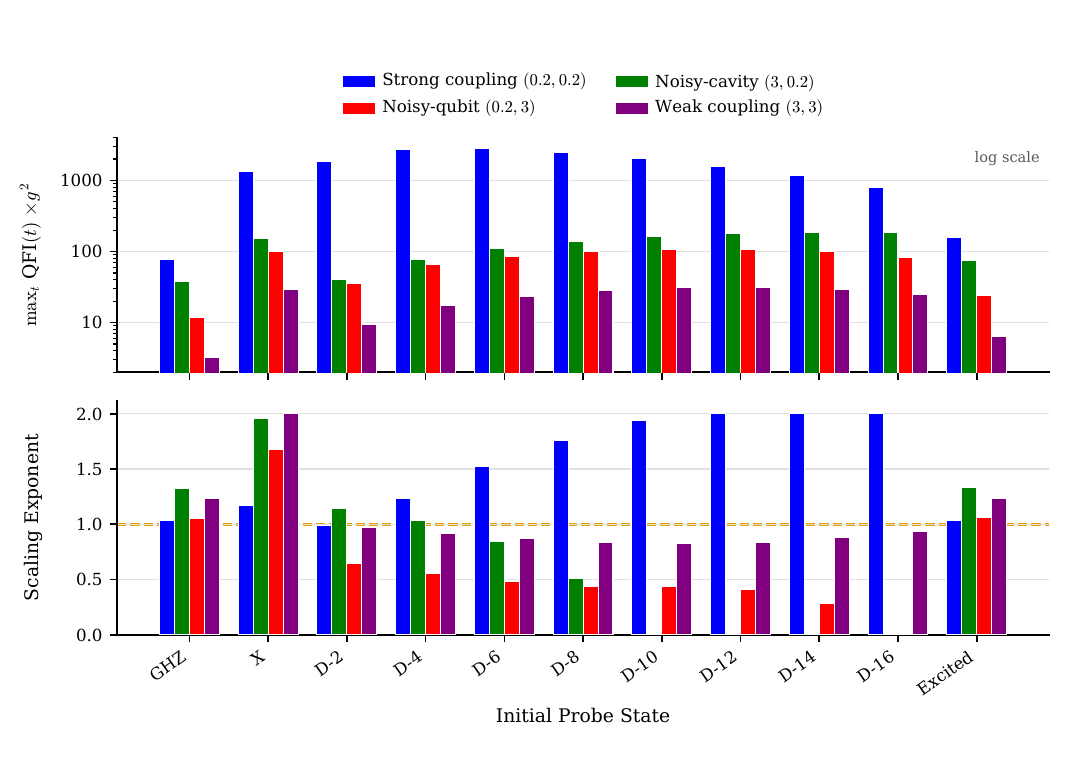}
        \put(3,58){\bfseries (a)}
        \put(3,33){\bfseries (b)}
    \end{overpic}
    \caption{
        Comparison of the metrological performance of the initial
        probe states in four representative coupling and loss regimes.
        (a) Maximum QFI at $N=20$, displayed on a logarithmic scale.
        (b) Finite-size scaling exponent $b$, obtained by fitting
        $\max_t \mathrm{QFI}(N)=aN^b+c$ over the available system sizes
        through $N=20$. The orange dashed line denotes the standard
        quantum limit, $b=1$. The regimes are strong coupling
        $(\kappa/g,\gamma/g)=(0.2,0.2)$, noisy-cavity
        $(3.0,0.2)$, noisy-qubit $(0.2,3.0)$, and weak coupling
        $(3.0,3.0)$. Here, D-$n$ denotes the Dicke state with $n$
        excitations, while ``Excited'' denotes the state in which all
        $N$ qubits are initially excited.
    }
    \label{fig:bar_plots}
\end{figure*}

Fig.~\ref{fig:bar_plots} presents a comparative analysis of the maximum QFIs and the scaling exponents across different states for all four regimes. In the context of the initially unentangled $X$-state, our bar plots in Fig.~\ref{fig:bar_plots} show an intriguing trend. For lower decay constant values, the $X$-state shows scaling closer to SQL, aligning with expectations. However, as we explore higher decay constant values, a notable shift towards scaling significantly better than SQL is observed. 


It is important to highlight that, while the $X$-state shows promising scaling properties, its maximum QFI, as revealed in our bar plots, is not always comparable to the magnitudes achievable with the Dicke states. In certain high-loss scenarios, the $X$-state's QFI approaches or matches that of the Dicke states, but there are regimes where Dicke states maintain a clear advantage in terms of QFI magnitude, particularly within the strong coupling regime. 

Saleem et al. showed that Dicke states can exceed SQL and even approach the Heisenberg limit under specific conditions involving low but non-negligible decay rates \cite{saleem2023achieving}. This finding is corroborated by the results shown in
Fig.~\ref{fig:bar_plots}(b). These results, however, are predominantly observed in the strong coupling limit where $g \gg \kappa, \gamma$, a regime that might be challenging to achieve in practical settings. In contrast, the X-state exhibits strongly superlinear finite-size scaling
in a regime of weaker coupling, characterized by $g \ll \kappa,\gamma$,
with fitted exponents approaching $b=2$ over the system sizes considered.
This suggests that the X-state could offer a promising route to enhanced
quantum metrological performance in environments where strong coupling
is not achievable.  

The favorable performance of the X-state in the high-loss regime can be
understood from the interplay between collective qubit--cavity coupling and
the absence of initial multipartite entanglement. Since
$|X\rangle=|+\rangle^{\otimes N}$, each qubit has maximal coherence between
its ground and excited states in the computational basis. The common cavity
mode converts these single-qubit coherences into collective sensitivity to
$g$, as reflected by the $N(N-1)$ cross terms in
$\langle J_+J_-\rangle$ discussed in Sec.~III. At the same time, unlike the
GHZ and Dicke probes, the X-state does not rely on initially prepared
multipartite entanglement, which is particularly vulnerable to dissipation \cite{Huelga1997,Escher2011}.
The parameter-dependent correlations relevant for sensing are instead
generated dynamically through the common cavity interaction \cite{benatti2003environment}. Consequently,
as the losses increase, the initially entangled probes are more strongly
degraded, while the collective coupling mechanism underlying the X-state's
sensitivity remains effective over the finite evolution time at which the
QFI is maximized. This explains why the X-state becomes comparatively
favorable in the high-loss regimes considered here.

Further analysis presented in Fig.~\ref{fig:bar_plots} reveals that the $X$-state surpasses other separable states in performance, particularly the Dicke-$0$ state (which is the ground state, $\ket{0}^{\otimes N}$) and the fully excited state $\ket{1}^{\otimes N}$, where $N$ denotes the total number of qubits comprising our probe. This finding indicates that the $X$-state's effectiveness in high-loss regimes is not merely a result of its separability. Instead, it points to unique characteristics of the $X$-state, likely tied to its initial configuration in the quantum state space, which enhance its performance under substantial losses. Additionally, the fully excited state shows comparable performance to the $X$-state in terms of the scaling exponent within the strong coupling and noisy-qubit regimes. However, our analysis indicates that, at least for decay rates ranging from $0.2$ to $3.0$, the maximum scaling exponent for the fully excited state does not surpass $1.5$. This contrasts with the $X$-state's ability to achieve the Heisenberg limit.

\section{Detuning estimation}
\label{sec:detune_est}
In our simulations focusing on detuning estimation, we set the detuning parameter $ \Delta = \omega_q - \omega_c $ to $\Delta$ = 0.1 $g$, by assigning $ \omega_q $ a value of 0.1 $g$ and keeping $ \omega_c $ at 0. 
This choice ensures that $ \Delta $ remains significantly smaller than $ g $, thus maintaining consistency in the system's dynamics. Similar to coupling estimation, when estimating detuning we are concerned with the QFI, except we take the partial derivative of the state $\rho$ with respect to $\Delta$ instead of $g$ in Eq. \ref{fisher1}. Across all four regimes, the $X$-state consistently outperformed other states in detuning estimation tasks, although its performance did not substantially exceed SQL. 

\begin{figure}
    \centering
    \includegraphics[width=0.8\columnwidth]{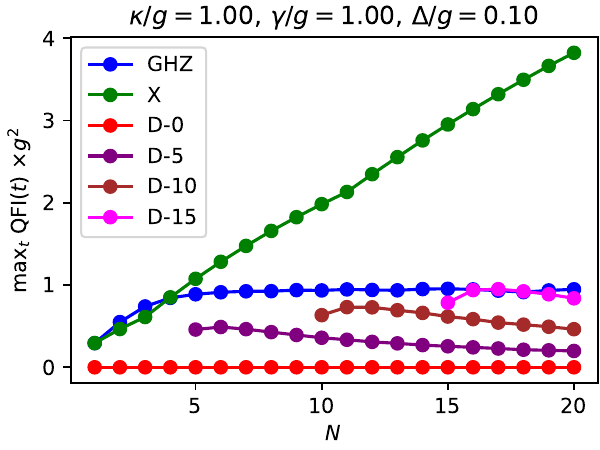}
    \caption{Scaling of maximum QFI against number of qubits $N$ when estimating detuning $\Delta$.}
    \label{fig:detune_plot}
\end{figure}

Interestingly, the maximum QFI values observed for detuning estimation were markedly lower than those for coupling estimation in all regimes. The highest maximum QFI values were recorded in the strong coupling regime, yet these did not surpass 10. For other states, including both GHZ and the Dicke states, the scaling exponent $ b $ hovered around 0. This result indicates that the GHZ and Dicke states are relatively ineffective for estimating detuning, in stark contrast to their performance in estimating the coupling strength. Example results in the moderate coupling regime $\gamma/g = \kappa/g = 1.0$ are shown in Figure~\ref{fig:detune_plot}. 

\section{Asymptotic Limit}

In Figure~\ref{fig:bar_plots}(b), it is evident that the scaling of the maximum QFI for the Dicke states in the strong coupling regime depends on the excitation number $n$. For instance, the scaling exponent reaches $2$ at $\kappa = \gamma = 0.2$ for $N = 1-20$, indicating HL scaling, while the scaling exponent decreases when $n$ falls below $10$. This raises the question of whether HL scaling still holds as the number of spins approaches infinity.

\begin{figure}
    \centering
    \begin{subfigure}{0.39\textwidth}
        \centering
        \includegraphics[width=\linewidth]{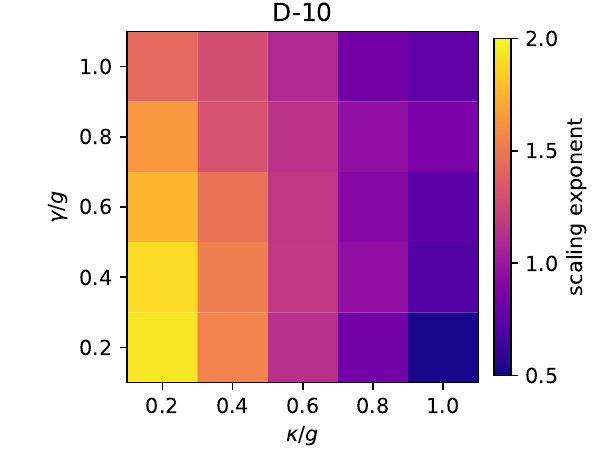}
        \label{fig:decay_sub1}
    \end{subfigure}\hfill
    \begin{subfigure}{0.39\textwidth}
        \centering
        \includegraphics[width=\linewidth]{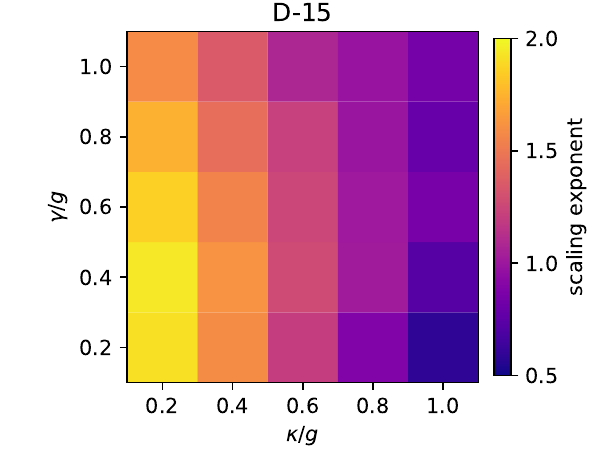}
        \label{fig:decay_sub2}
    \end{subfigure}\hfill
    \begin{subfigure}{0.39\textwidth}
        \centering
        \includegraphics[width=\linewidth]{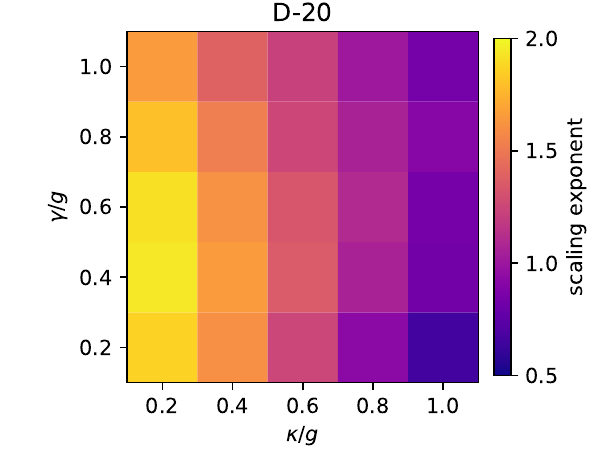}
        \label{fig:decay_sub3}
    \end{subfigure}
    \caption{Scaling plots for the Dicke states with the number of excitations set to half the total number of qubits. The top panel shows the scaling exponent plot for $10-20$ qubits, the middle panel for $15-30$ qubits, and the lower panel for $20-40$ qubits.}
    \label{fig:decay_exponent_plots}
\end{figure}

Initially, we found that for constant $n$, the scaling becomes gradually more linear as the number of spins increases beyond $N = 20$. However, we found that the critical factor for evaluating the precision scaling of the Dicke states is the ratio between the excitation number $n$ and the total number of spins $N$. Our results suggest that HL scaling is preserved as $N$ increases to infinity, provided that the ratio $n/N$ remains constant. In Figure \ref{fig:decay_exponent_plots}, we present scaling plots for the Dicke states with $n/N = 1/2$ within the strong coupling regime. Intuitively, the dependence on the ratio $n/N$ should make sense, as the sensing capabilities will depend to some extent on the initial degrees of entanglement, and it has been shown that the geometric measure of entanglement is maximized for Dicke states at around $n/N = 1/2$ \cite{Martin_2010}.

Similar to Figure~\ref{fig:bar_plots}, we fit the curve $y(N) = a N^b + c$ to the maximum QFI data points and plot the exponent $b$ against the decay rates $\kappa$ and $\gamma$. For the Dicke-$10$ state, the fit is applied to data points where $N$ ranges from $10-20$; for the Dicke-$15$ state, $N = 15-30$; and for the Dicke-$20$ state, $N = 20-40$. We find that the maximum scaling exponent remains around $1.8$ at $\kappa = \gamma = 0.2$ across the Dicke states. Similar values for the scaling exponent are observed across the different Dicke states for all other decay rates within the strong coupling regime.

\section{Dephasing on Qubits}


In this section, we consider a variation of the qubits-cavity model where the qubits are affected by dephasing. We allow $\kappa$ to still represent the decay rate of the cavity, but $\gamma$ now denotes the dephasing rate acting on the qubits. This change appears in the third term of Equation~\ref{eq:Lindblad}, where $\sigma_z^{(j)}$ replaces $\sigma_-^{(j)}$:

\begin{equation}
\dot{\rho} = -i[H, \rho] + \kappa D[a](\rho) + \gamma \sum_{j=1}^{N} D[\sigma_z^{(j)}](\rho).
\label{eq:Lindblad_dephase}
\end{equation}

In this formulation, $\sigma_z^{(j)}$ represents the Pauli-$Z$ operator acting on the $j$-th qubit. Dephasing, unlike decay, preserves the energy of the qubits but randomizes their phases, leading to a loss of coherence without energy dissipation.

Given that the polarization of the dephasing operator $\sigma_z^{(j)}$ differs from the decay operator $\sigma_-^{(j)}$, we decided to explore new states, specifically the Dicke-$X$ and GHZ-$X$ states. The Dicke-$X$ state is a variation of the Dicke states, except it forms a product state of $\ket{+}$ and $\ket{-}$, with $n$ representing the number of spins in the $\ket{+}$ state. Mathematically, a Dicke-$X$-$n$ state can be expressed as:

\begin{equation}
|\text{Dicke-X-n}\rangle = \frac{1}{\sqrt{\binom{N}{n}}} \sum_{\text{permutations}} \left|\underbrace{+\cdots+}_{n}\underbrace{-\cdots-}_{N-n}\right\rangle.
\end{equation}

Similarly, the GHZ-X state is an equal superposition of the $|+\rangle$ and $|-\rangle$ states, where $|\pm\rangle = (|0\rangle \pm |1\rangle)/\sqrt{2}$ and $\sigma_x|\pm\rangle = \pm|\pm\rangle$. Analogous to the conventional GHZ state expressed in the computational basis, the GHZ-X state is given by:

\begin{equation}
|\text{GHZ-X}\rangle = \frac{1}{\sqrt{2}} \left( \ket{+}^{\otimes N} + \ket{-}^{\otimes N} \right).
\end{equation}

Under the Lindblad evolution described by Equation~\ref{eq:Lindblad_dephase}, we found that the Dicke-$X$ and GHZ-$X$ states exhibit behavior similar to the $X$-polarized state's behavior in the original qubit decay case. Like the $X$-polarized state, the Dicke-$X$ and GHZ-$X$ states achieve the best precision scaling under moderate to weak coupling regimes, approaching or achieving HL scaling within $N=1-20$ qubits.

\begin{figure}[t]
    \centering
    \includegraphics[width=0.4\textwidth]{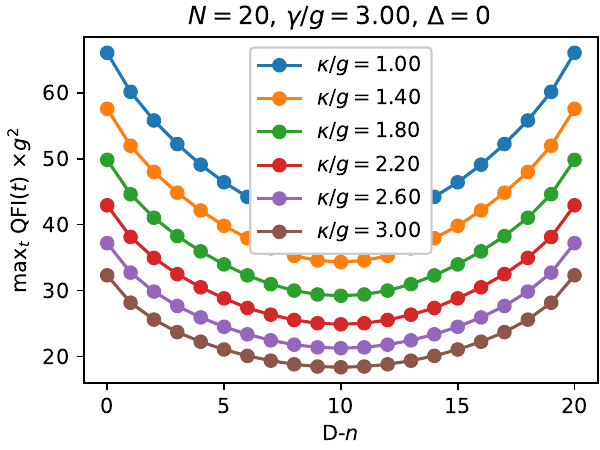} 
    \caption{Maximum QFI for the Dicke-$X$-$n$ initial state as a function of the number $n$ of qubits in the $\ket{+}$ state, with fixed qubit decay rate $\gamma/g=3.0$ and varying cavity decay rate $\kappa/g$.}
    \label{fig:Dicke-x_plots_coupling_dephase}
\end{figure}

Another interesting finding is that the Dicke-$X$ states, for all regimes, has maximum QFI values at the end points $\ket{+}^{\otimes N}$ and $\ket{-}^{\otimes N}$, and reaches its minimum at $n/N = 1/2$, as illustrated in Figure~\ref{fig:Dicke-x_plots_coupling_dephase}. The state where $n/N = 1/2$ for the Dicke-$X$ state corresponds to the GHZ state, which, as pointed out by \cite{saleem2023achieving}, lacks robustness to noise and is therefore a poor initial probe state. Additionally, while the QFI magnitudes decrease, the precision scaling slightly increases as $n$ increases from $n=0$ to $n=N/2$. The GHZ-$X$ state also exhibits similar scaling and magnitudes to the Dicke-$X$-$0$ state (equivalent to the $X$-state).

\section{Conclusions and future directions}
\label{sec:conclusion}
We simulated a qubits-cavity system with up to $N=20$ qubits to evaluate the metrology performance for various probe states, primarily focusing on high-loss regimes, such as noisy-cavity, noisy-qubit, and weak coupling conditions. Remarkably, we discovered that the fully separable $X$-polarized state not only exceeds the standard quantum limit but often reaches the Heisenberg limit in these scenarios. This highlights the $X$-state's adaptability and noise resilience, making it a promising candidate for practical quantum metrology applications in environments characterized by significant losses.

One of the overarching themes of our research is the balance between entanglement and noise resilience. Unlike highly entangled states, such as the GHZ state which have diminished performance in noisy settings, the $X$-state shows a robustness against noise, offering better performance in high-loss conditions. This finding suggests that in certain practical quantum metrology scenarios, especially those involving substantial losses, states with no initial entanglement might be more advantageous. Future research will involve understanding the unique properties of the $X$-state that contribute to its good performance in noisy settings that not only set it apart from entangled states like the GHZ and Dicke states but also other separable states.

Solid-state spin qubits provide a potential platform for realizing the
qubit--cavity sensing scenarios considered here. Semiconductor spin qubits
have been coupled to superconducting microwave resonators over a broad range
of coupling and dissipation rates
\cite{Petersson2012,Ungerer2024}. For example, Petersson \textit{et al.}
reported a superconducting resonator with
$f_c\simeq6.2~\mathrm{GHz}$ and $Q\simeq2000$, corresponding to
$\kappa/2\pi\simeq3.1~\mathrm{MHz}$, and estimated a spin--cavity coupling
$g/2\pi\simeq0.2~\mathrm{MHz}$, with coupling rates approaching
$1~\mathrm{MHz}$ considered experimentally feasible
\cite{Petersson2012}. These values correspond to
$\kappa/g\simeq15.5$ for the reported coupling and
$\kappa/g\simeq3.1$ for $g/2\pi\simeq1~\mathrm{MHz}$.

More recently, strong spin--photon coupling was experimentally demonstrated
in an InAs nanowire double quantum dot, with
$g/2\pi=139~\mathrm{MHz}$,
$\kappa/2\pi=19.8~\mathrm{MHz}$, and
$\gamma/2\pi=116~\mathrm{MHz}$, corresponding to
$\kappa/g\simeq0.14$ and
$\gamma/g\simeq0.83$
\cite{Ungerer2024}. Thus, experimentally accessible solid-state cQED
systems span parameter regimes comparable to both the strong-coupling
and loss-dominated regimes investigated in the present work.

 
Acknowledgements:  This material is based upon work supported by the U.S. Department of Energy Office of Science National Quantum Information Science Research Centers. Q.L., S.K.G. and Z.H.S. were supported by the Q-NEXT Center. Work performed at the Center for Nanoscale Materials, a U.S. Department of Energy Office of Science User Facility, was supported by the U.S. DOE Office of Basic Energy Sciences, under Contract No. DE-AC02-06CH11357. A.~S.~was supported by QuEST grant No Q-113 of the Department of Science and Technology, Government of India. The results in this work were obtained from simulations using $\sim$1000 CPU-hours on the Bebop computing cluster at Argonne National Laboratory.

\bibliography{references}

@article{cramer1946contribution,
  title={A contribution to the theory of statistical estimation},
  author={Cram{\'e}r, Harald},
  journal={Scandinavian {A}ctuarial {J}ournal},
  volume={1946},
  number={1},
  pages={85--94},
  year={1946},
  publisher={Taylor \& Francis}
}

@article{braunstein1994statistical,
  title={Statistical distance and the geometry of quantum states},
  author={Braunstein, Samuel L and Caves, Carlton M},
  journal={Physical {R}eview {L}etters},
  volume={72},
  number={22},
  pages={3439},
  year={1994},
  publisher={APS}
}

@article{liu2019quantum,
  title={Quantum Fisher information matrix and multiparameter estimation},
  author={Liu, Jing and Yuan, Haidong and Lu, Xiao-Ming and Wang, Xiaoguang},
  journal={Journal of {P}hysics {A}: {M}athematical and {T}heoretical},
  volume={53},
  number={2},
  pages={023001},
  year={2019},
  publisher={IOP Publishing}
}

@article{lindblad_generators_1976,
	title = {On the generators of quantum dynamical semigroups},
	volume = {48},
	issn = {1432-0916},
	url = {https://doi.org/10.1007/BF01608499},
	doi = {10.1007/BF01608499},
	number = {2},
	urldate = {2022-08-28},
	journal = {Communications in {M}athematical {P}hysics},
	author = {Lindblad, G.},
	month = jun,
	year = {1976},
	pages = {119--130}
}

@article{chruscinski_brief_2017,
	title = {A {Brief} {History} of the {GKLS} {Equation}},
	volume = {24},
	issn = {1230-1612},
	url = {https://www.worldscientific.com/doi/abs/10.1142/S1230161217400017},
	doi = {10.1142/S1230161217400017},
	number = {03},
	urldate = {2022-08-28},
	journal = {Open {S}ystems \& {I}nformation {D}ynamics},
	author = {Chruściński, Dariusz and Pascazio, Saverio},
	month = sep,
	year = {2017},
	pages = {1740001}
}

@article{polino2020photonic,
  title={Photonic quantum metrology},
  author={Polino, Emanuele and Valeri, Mauro and Spagnolo, Nicol{\`o} and Sciarrino, Fabio},
  journal={{AVS} {Q}uantum {S}cience},
  volume={2},
  number={2},
  pages={024703},
  year={2020},
  publisher={American Vacuum Society}
}

@article{lu2010quantum,
  title={Quantum Fisher information flow and non-{M}arkovian processes of open systems},
  author={Lu, Xiao-Ming and Wang, Xiaoguang and Sun, CP},
  journal={Physical {R}eview {A}},
  volume={82},
  number={4},
  pages={042103},
  year={2010},
  publisher={APS}
}

@article{gammelmark2014fisher,
  title={Fisher information and the quantum Cram{\'e}r-Rao sensitivity limit of continuous measurements},
  author={Gammelmark, S{\o}ren and M{\o}lmer, Klaus},
  journal={Physical {R}eview {L}etters},
  volume={112},
  number={17},
  pages={170401},
  year={2014},
  publisher={APS}
}

@article{altintas2016quantum,
  title={Quantum Fisher information of an open and noisy system in the steady state},
  author={Altintas, Azmi Ali},
  journal={Annals of {P}hysics},
  volume={367},
  pages={192--198},
  year={2016},
  publisher={Elsevier}
}

@article{saleem2022optimal,
  title={Optimal time for sensing in open quantum systems},
  author={Saleem, Zain H and Shaji, Anil and Gray, Stephen K},
  journal={Physical {R}eview {A}},
  volume={108},
   pages={022413},
  year={2023}
}

@article{giovannetti2004quantum,
  title={Quantum-enhanced measurements: beating the standard quantum limit},
  author={Giovannetti, Vittorio and Lloyd, Seth and Maccone, Lorenzo},
  journal={Science},
  volume={306},
  number={5700},
  pages={1330--1336},
  year={2004},
  publisher={American Association for the Advancement of Science}
}

@article{chase2008collective,
  title={Collective processes of an ensemble of spin-1/2 particles},
  author={Chase, Bradley A and Geremia, JM},
  journal={Physical {R}eview {A}},
  volume={78},
  number={5},
  pages={052101},
  year={2008},
  publisher={APS}
}

@article{shammah2018open,
  title={Open quantum systems with local and collective incoherent processes: Efficient numerical simulations using permutational invariance},
  author={Shammah, Nathan and Ahmed, Shahnawaz and Lambert, Neill and De Liberato, Simone and Nori, Franco},
  journal={Physical {R}eview {A}},
  volume={98},
  number={6},
  pages={063815},
  year={2018},
  publisher={APS}
}

@article{toth2014quantum,
  title={Quantum metrology from a quantum information science perspective},
  author={T{\'o}th, G{\'e}za and Apellaniz, Iagoba},
  journal={Journal of {P}hysics {A}: {M}athematical and {T}heoretical},
  volume={47},
  number={42},
  pages={424006},
  year={2014},
  publisher={IOP Publishing}
}

@book{nawrocki2015introduction,
  title={Introduction to quantum metrology},
  edition={Second},
  author={Nawrocki, Waldemar},
  year={2019},
  publisher={Springer Nature Switzerland},
  address={Cham, Switzerland}
}

@article{benatti2003environment,
  title={Environment induced entanglement in {M}arkovian dissipative dynamics},
  author={Benatti, Fabio and Floreanini, Roberto and Piani, Marco},
  journal={Physical {R}eview {L}etters},
  volume={91},
  number={7},
  pages={070402},
  year={2003},
  publisher={APS}
}

@article{gorini_completely_1976,
	title = {Completely positive dynamical semigroups of {N}‐level systems},
	volume = {17},
	issn = {0022-2488},
	url = {https://aip.scitation.org/doi/abs/10.1063/1.522979},
	doi = {10.1063/1.522979},
	number = {5},
	urldate = {2022-08-28},
	journal = {{Journal of Mathematical Physics}},
	author = {Gorini, Vittorio and Kossakowski, Andrzej and Sudarshan, E. C. G.},
	month = may,
	year = {1976},
	note = {Publisher: American Institute of Physics},
	pages = {821--825}
}

@article{manzano2020short,
  title={A short introduction to the {L}indblad master equation},
  author={Manzano, Daniel},
  journal={{AIP} Advances},
  volume={10},
  number={2},
  pages={025106},
  year={2020},
}

@Inbook{Greenberger1989,
author="Greenberger, Daniel M.
and Horne, Michael A.
and Zeilinger, Anton",
editor="Kafatos, Menas",
title="Going Beyond {B}ell's Theorem",
bookTitle="Bell's Theorem, Quantum Theory and Conceptions of the Universe",
year="1989",
publisher="Springer Netherlands",
address="Dordrecht",
pages="69--72",
}

@misc{saleem2023achieving,
      title={Achieving the Heisenberg limit with Dicke states in noisy quantum metrology}, 
      author={Zain H. Saleem and Michael Perlin and Anil Shaji and Stephen K. Gray},
      year={2023},
      eprint={2309.12411},
      archivePrefix={arXiv},
      primaryClass={quant-ph}
}

@inproceedings{Faist:21,
author = {Philippe Faist and Mischa P. Woods and Victor V. Albert and Joseph M. Renes and Jens Eisert and John Preskill},
booktitle = {Quantum Information and Measurement VI 2021},
journal = {Quantum Information and Measurement VI 2021},
pages = {W2A.3},
publisher = {Optica Publishing Group},
title = {Time-energy uncertainty relation for noisy quantum metrology},
year = {2021},
url = {https://opg.optica.org/abstract.cfm?URI=QIM-2021-W2A.3},
doi = {10.1364/QIM.2021.W2A.3},
}

@article{Akipour2014Quantum,
  title = {Quantum Metrology in Open Systems: Dissipative Cram\'er-Rao Bound},
  author = {Alipour, S. and Mehboudi, M. and Rezakhani, A. T.},
  journal = {Phys. Rev. Lett.},
  volume = {112},
  issue = {12},
  pages = {120405},
  numpages = {6},
  year = {2014},
  month = {Mar},
  publisher = {American Physical Society},
  doi = {10.1103/PhysRevLett.112.120405},
  url = {https://link.aps.org/doi/10.1103/PhysRevLett.112.120405}
}

@Article{Naghiloo2019,
author={Naghiloo, M.
and Abbasi, M.
and Joglekar, Yogesh N.
and Murch, K. W.},
title={Quantum state tomography across the exceptional point in a single dissipative qubit},
journal={Nature Physics},
year={2019},
month={Dec},
day={01},
volume={15},
number={12},
pages={1232-1236},
issn={1745-2481},
doi={10.1038/s41567-019-0652-z},
url={https://doi.org/10.1038/s41567-019-0652-z}
}

@article{Martin_2010,
   title={Multiqubit symmetric states with high geometric entanglement},
   volume={81},
   ISSN={1094-1622},
   url={http://dx.doi.org/10.1103/PhysRevA.81.062347},
   DOI={10.1103/physreva.81.062347},
   number={6},
   journal={Physical Review A},
   publisher={American Physical Society (APS)},
   author={Martin, J. and Giraud, O. and Braun, P. A. and Braun, D. and Bastin, T.},
   year={2010},
   month=jun }

@article{Petersson2012,
  author  = {Petersson, K. D. and McFaul, L. W. and Schroer, M. D. and
             Jung, M. and Taylor, J. M. and Houck, A. A. and Petta, J. R.},
  title   = {Circuit quantum electrodynamics with a spin qubit},
  journal = {Nature},
  volume  = {490},
  pages   = {380--383},
  year    = {2012}
}

@article{Ungerer2024,
  author  = {Ungerer, J. H. and Pally, A. and Kononov, A. and Lehmann, S. and
             Ridderbos, J. and Potts, P. P. and Thelander, C. and Dick, K. A. and
             Maisi, V. F. and Scarlino, P. and Baumgartner, A. and
             Sch{\"o}nenberger, C.},
  title   = {Strong coupling between a microwave photon and a singlet-triplet qubit},
  journal = {Nature Communications},
  volume  = {15},
  pages   = {1068},
  year    = {2024}
}

@article{Huelga1997,
  title = {Improvement of Frequency Standards with Quantum Entanglement},
  author = {Huelga, S. F. and Macchiavello, C. and Pellizzari, T. and
            Ekert, A. K. and Plenio, M. B. and Cirac, J. I.},
  journal = {Physical Review Letters},
  volume = {79},
  pages = {3865--3868},
  year = {1997},
  doi = {10.1103/PhysRevLett.79.3865}
}

@article{Escher2011,
  title = {General Framework for Estimating the Ultimate Precision Limit
           in Noisy Quantum-Enhanced Metrology},
  author = {Escher, B. M. and de Matos Filho, R. L. and Davidovich, L.},
  journal = {Nature Physics},
  volume = {7},
  pages = {406--411},
  year = {2011},
  doi = {10.1038/nphys1958}
}
\pagebreak

The submitted manuscript has been created by UChicago Argonne, LLC, Operator of Argonne National Laboratory (``Argonne''). Argonne, a U.S. Department of Energy Office of Science laboratory, is operated under Contract No. DE-AC02-06CH11357. The U.S. Government retains for itself, and others acting on its behalf, a paid-up nonexclusive, irrevocable worldwide license in said article to reproduce, prepare derivative works, distribute copies to the public, and perform publicly and display publicly, by or on behalf of the Government. The Department of Energy will provide public access to these results of federally sponsored research in accordance with the DOE Public Access Plan (\url{http://energy.gov/downloads/doe-public-access-plan}).

\appendix
\section{Quantum Fisher Information}
\label{app:qfi}

Consider a quantum state described by the density matrix $\rho(\theta)$, which depends on the parameter $\theta$. The QFI, $F(\theta)$, is given by \cite{liu2019quantum}:
\begin{align}
    \label{fisher1}
F(\theta) = \sum_{\{a,b|\lambda_a+\lambda_b \neq 0\}} \frac{2 }{\lambda_{a}+\lambda_{b}}|\langle\lambda_{a}|\partial_{\theta}\rho |\lambda_{b}\rangle|^2~~~.
\end{align}
In our work, $\theta$ is taken to be either the qubit-cavity coupling, $g$, or the difference between qubit and cavity transition energies, $\Delta$. 

In the context of quantum metrology, QFI is used to bound the precision of parameter estimation. According to the Quantum Cramér-Rao Bound \cite{cramer1946contribution}, the variance of any unbiased estimator $\hat{\theta}$ of the parameter $\theta$ is bounded by the inverse of the QFI:
\begin{equation}
    \text{Var}(\hat{\theta}) \geq \frac{1}{M F(\theta)}~~,
\end{equation}

\noindent where $M$ corresponds to the number of independent experiments
carried out.  The square root of the variance corresponds to the uncertainty in the parameter.  This bound implies that higher QFI values allows for more precise parameter estimation.

\end{document}